\documentstyle[12pt]{article}

\begin{document}

\Large
\bf

\begin{center}
SPIN CONTENT OF THE QUANTUM SOLITON
\end{center}

\begin{center}
A.Dubikovsky and K.Sveshnikov\footnote{\large E-mail: costa@bog.msu.su}
\end{center}

\normalsize
\it

\begin{center}
Department of Physics, Moscow State University, Moscow 119899, Russia
\end{center}

\leftskip 1 true cm
\rightskip 1 true cm
\baselineskip 10 pt

\vskip 0.5 true cm
\normalsize
\rm

The classical soliton solution, quantized by means of suitable
translational and rotational collective  coordinates, is embedded into the
one-particle irreductible representation of the Poincare group
corresponding to a definite spin.  It is shown, that within the
conventional quasiclassical expansion such embedding leads to a set of
nontrivial consistency conditions imposed on the classical solution. The
validity of these relations is considered for a number of soliton models
in 2+1- and 3+1-dimensions.

\leftskip 0 true cm
\rightskip 0 true cm
\baselineskip 11 pt
\vskip 0.5 true cm

\large
\rm

Solitons with spin play an essential role in describing extended
particles, such as baryons in the soliton skyrmeon models in
3+1-dimensional spacetime \cite{skyrme}, and anyons with fractional spin
and statistics in the planar case \cite{anyone}. Within the framework of
the quasiclassical soliton quantization it is  assumed usually,
that the spherically non-symmetric classical soliton
solution acquires the spin by introducing suitable rotational collective
coordinates \cite{rajaraman}. Such procedure restores the broken
rotational symmetry by cancellation of corresponding zero modes and yields
states with physical quantum  numbers of the total angular momentum
operators. In the planar case, such procedure has been studied in
ref. \cite{umezawa}.  In 3+1-dimensions a typical example of such kind gives
the recent development of skyrmeon baryon model (see refs. \cite{skyrmeon}
and references therein).  However, in the lorentz-covariant theory the
consistent interpretation of the resulting ground state as an extended
particle with a spin should be more restrictive. Namely, it requires the
fulfillment of the one-particle irreductible representation of the Poincare
group corresponding to the same spin  and mass \cite{barut}.

In the present paper, we give an  analysis of such embedding for the
soliton, quantized by means of translational and rotational collective
coordinates, into corresponding representation of the Poincare group. The
result is that the consistency of this embedding is provided by an
additional set of nontrivial integral relations imposed on the classical
soliton solution.  These conditions do not be automatically consistent
with the equations of motion and so yield some superselection rules for
the classical solution to be interpreted as a spinning particle upon
quantization. We show also that the typical hedgehog-type classical
configurations of the chiral $\sigma$-models satisfy these conditions
independently of the choice of the shape of the chiral angle, both in 2+1-
and 3+1-spacetimes.

As a first step, we consider a nonlinear scalar field in
$3$ spatial dimensions, described by the Lagrangean density \equation
   {\cal L} = {1 \over 2}{(\partial_{\mu} \varphi)}^{2}- U(\varphi),
\label{lagr}
\endequation
which possesses a static soliton solution \equation
\varphi_{c}(x)=u(\vec x). \label{sol} \endequation
 According to the virial theorem,
such solutions are  unstable in more then one spatial dimension,
but for our purposes it is not so important compared to
simplicity of presentation. The angular momenta $ J^{i} $ and Lorentz
boosts $ K^{i} $ are given by \eqnarray
   J^{i} & = & \int d \vec x \; \varepsilon_{ijk}\;x^{j}\;T^{k0} =\int d
       \vec x\;\varepsilon_{ijk}\;x_{j}\;\partial_{k}\varphi \;\pi.
\label{moment}
\\
      K^{i} & = & \int d \vec x \; \left( x^{0}T^{i0}-x^{i}T^{00} \right)
	    =x^{0}P^{i}-\int d \vec x \; x^{i}\; {\cal H},
\label{boost}
\endeqnarray
   In these expressions $ \;T^{\mu \nu}\; $ is the energy--momentum tensor,
$ \;P^{i}=\int d \vec x \; T^{i0} $ are the spatial momenta,
$ \;{\cal H}=T^{00}=\pi \dot \varphi - {\cal L} (\varphi) $ is the Hamiltonian
density
and
$ \;\pi=\partial {\cal L} / \partial \dot \varphi $ is the canonical field
momentum.

Quantization of the soliton sector in the neighborhood of the classical
solution (\ref{sol}) proceeds
within the canonical collective coordinate framework \cite{collcoord}. For
our purposes it is convenient to consider the field $\varphi(\vec x)$ in
the
$\hbox{\rm Schr}{\ddot {\hbox{\rm o}}} \hbox{\rm dinger}$
picture. The
substitution, introducing translational and rotational collective
coordinates, reads
\equation \varphi(\vec x) = u\left( R^{-1}(\vec c)(\vec
  x-\vec q)\right) +\Phi\left( R^{-1}(\vec c)(\vec x-\vec q)\right),
\label{17}
\endequation
where $\Phi$ is the meson field, $R(\vec c)$ is the rotation matrix,
$\vec q$ and $\vec c$ are the translational and rotational
collective coordinates correspondingly.

For  the rotation group we use the vector-parametrization  \cite{Fedorov},
when the rotation matrix $R(\vec c)$ is represented as
\equation
   R(\vec c)=1+2 \; {c^{\times}+{c^{\times}}^{2} \over 1+c^{2}}
    ={1-c^{2}+2c^{\times}+2c\cdot c \over 1+c^{2}},
\label{18}
\endequation
where $c^{\times}_{ab}=\varepsilon_{adb} c_{d}$,
${(c\cdot c)}_{ab}=c_{a}c_{b}$, $c^{2}=\vec c \; \vec c.$
The composition law for vector-parameters, corresponding to product
of rotations $R(\vec a) \; R(\vec b)=R(\vec c),$
is given by
\equation
    \vec c=\langle \vec a, \; \vec b \rangle={\vec a+\vec b+\vec a \times
               \vec b \over 1-\vec a \; \vec b},
\label{19}
\endequation
what is the simplest group composition law after the abelian one
\cite{Fedorov}.  The generators of infinitesimal rotations are \equation
    \vec S=-{i \over 2} \left( 1+c\cdot c +c^{\times} \right)
          {\partial \over  \partial \vec c} \ ,
\label{20}
\endequation
while the finite rotations $U(\vec a)$, defined so that $U^{+}(\vec a)
\; \vec b \; U(\vec a)=\langle \vec a, \; \vec b \rangle$, take the form
\equation
   U(\vec a)= \exp\left\{-2i\vec a\vec S\right\}.
\label{21}
\endequation

Returning to the decomposition (\ref{17}), one finds that the
total momentum of the field is now represented as
\equation
   \vec P=-i{\partial \over \partial \vec q} \ ,
\label{22}
\endequation
and the total angular momentum is equal to
\equation
   \vec J=\vec L+\vec S,
\label{23}
\endequation
where $\vec L=\vec q \times \vec P$ is the orbital
momentum and the spin $\vec S$ is defined by eq.(\ref{20}).

The gauge fixing conditions, preserving the total number of degrees
of freedom, are chosen in the conventional form
as linear constraints imposed on the meson field $\Phi(\vec y)$
\cite{collcoord}
\equation
  \int d \vec y \; N^{(\alpha)}(\vec y) \; \Phi(\vec y)=0.
\label{24}
\endequation
The set  $\{ N^{(\alpha)}(\vec y) \}$ should ensure the cancellation
of the zero-frequency modes from the mesonic spectrum and is defined as
follows.
In the general case, the spherically non-symmetric classical solution
$u(\vec x)$ yields three
translational zero modes \equation
 \psi_{i}(\vec x)=\partial_{i}u(\vec x) \label{trans},
\endequation
and three rotational \equation
f_{i}(\vec x)=\varepsilon_{ijk}x_{j}\partial_{k}u(\vec x). \label{rotat}
\endequation
Let us denote $M^{(\alpha)}(\vec y) = \{ \psi_{i}(\vec y), \;
f_{i}(\vec y) \}.$
Then $N^{(\alpha)}(\vec y)$
are given by linear combinations of $M^{(\beta)}(\vec y)$ subject of
relations
\equation
  \int d \vec y \; N^{(\alpha)}(\vec y) \; M^{(\beta)}(\vec y)=
   \delta_{\alpha \beta}.
\label{25}
\endequation
The crucial point here is that, under very general conditions in the
lorentz-covariant field theory described by the Lagrangean (\ref{lagr})
the system of zero-frequency modes is orthogonal \cite{paper}, namely
\eqnarray
    \int d \vec \xi \; \psi_{i}(\vec \xi) \;
      \psi_{j}(\vec \xi) & = & M\delta_{ij},
\label{ortpp}
\\
     \int d \vec \xi \; \psi_{i}(\vec \xi) \; f_{j}(\vec \xi)  & = & 0,
\label{ortpf}
\\
     \int d \vec \xi \; f_{i}(\vec \xi) \; f_{j}(\vec \xi) & = &
               \Omega_{ij}  =  \Omega_{i}\delta_{ij} \ ,
\label{ortff}
\endeqnarray
where $M$ is the mass and $\Omega_{ij}$ are the moments of inertia
of the classical solution, so one immediately gets
\equation
  N^{(\alpha)}(\vec y)=\{\psi_{i}(\vec y)/M, \; f_{i}(\vec y)/\Omega_{i}\}.
\label{26}
\endequation

Now let us briefly discuss the corresponding transformation
 of the canonical momentum $\pi(\vec x)$. In the
$\hbox{\rm Schr}{\ddot {\hbox{\rm o}}}\hbox{\rm dinger}$
picture, it is
calculated as a composite derivative
\eqnarray
\pi(\vec x)=-i{\delta
\over \delta \varphi(\vec x)} & = & {\partial \vec c \over \partial \varphi
     (\vec x)} \left( -i{\partial \over \partial \vec c } \right)
\\
\nonumber
     & + & {\partial \vec q \over \partial \varphi(\vec x)} \left(
   -i{\partial \over \partial \vec q} \right) +\int d \vec y \;
   {\delta\Phi(\vec y) \over \delta\varphi(\vec x)} \left( -i{\delta \over
   \delta\Phi(\vec y)} \right).
\label{27}
\endeqnarray
Here it should be noted, that in this approach the resulting expression
for $\pi(\vec x)$ is not explicitly hermitian, since one has to take into
account the change in the functional measure after introducing the
collective variables \cite{parmentola}.  However,
these effects are of essentially quantum origin and therefore lye beyond
the leading quasiclassical approximation, which will be used below.

Now,
calculating from  eq.(\ref{24}) the derivatives $ \partial \vec q /
\partial \varphi(\vec x) $ and $ \partial \vec c / \partial \varphi(\vec
x)$  and taking account of the orthogonality conditions
(\ref{ortpp}-\ref{ortff}), one gets
the following lowest-order contribution to $\pi(\vec x)$ from collective
degrees of freedom (for details see ref.\cite{paper}) \equation \pi(\vec
    x) = - \psi_{i} (\vec \xi) \; R^{-1}_{ij}(\vec c) \; { P_{j} \over M}
    - f_{i} (\vec \xi) \; \Omega^{-1}_{ik} \; R^{-1}_{kl}(\vec c) \;
   S_{l}, \label{impuls} \endequation In eq.(\ref{impuls}) $ \vec \xi =
   R^{-1}(\vec c)(\vec x-\vec q) $, $ \vec P$ and $ \vec S $  are the
momentum and the spin of the field defined earlier, and the contributions
to $\pi(\vec x)$ from the meson field and momentum are completely neglected. So
the expression (\ref{impuls}) should be identified indeed with the
 canonical momentum of the soliton field in the leading quasiclassical
approximation.

Now we put the quasiclassical soliton field $u(R^{-1}(\vec x - \vec
q))$ and its canonical momentum defined in eq.(\ref{impuls}) into
corresponding Noether expressions for Lorentz generators (3,4) and demand
for their coincidence with the corresponding one-particle representation
of the Poincare group with the same mass $M$ and spin $S$. It means, that
the Lorentz generators $J^{\mu \nu}$ should take the form \cite{barut}
\eqnarray
   J^{i} & = & \varepsilon_{ijk}\;q^{j}P^{k} +S^{i} ,
\label{momentrep}
   \\
   K^{i} & = & q^{0}P^{i}-q^{i}P^{0}
		-{\varepsilon_{ijk}\;P^{j}S^{k} \over P^{0}+M}.
\label{boostrep}
\endeqnarray

   Firstly, it is a trivial task to verify, that inserting into eq.(3) the
soliton operators, one gets identically the eq.(\ref{momentrep}), provided
by the orthogonality conditions (\ref{ortpp}--\ref{ortff}). Actually,
this result reflects the relation (\ref{23}) for the total angular momentum,
which is the direct consequence of the group properties included into
the initial eq.(\ref{17}) and therefore holds for all orders of
perturbation expansion.

Applying the same procedure to the Lorentz boost operators (\ref{boost}),
in the first step we note,
that the term
$$
   \int d \xi \; \xi^{i} \; \left( \; {1 \over 2} {(\vec
  \partial u)}^{2} + U(u) \right) (\vec \xi)
$$
can be always cancelled by a spatial translation, what means that
the center of mass of the classical soliton is fixed at the origin
in the $\vec \xi$-reference frame.  So we obtain
\equation
  K^{i}=x^{0}P^{i}-q^{i}P^{0}-\int d \vec \xi \;R^{ij}\xi_{j} \; {1 \over
  2}  \pi^{2}(\vec \xi) .
\label{ppp}
\endequation
Inserting the expression (21) into eq.(24), we get further
\eqnarray K^{i} & = &
x^{0}P^{i}-q^{i}P^{0}- \nonumber \\ & &
-(\Omega^{-1}_{kl}R^{-1}_{lm}S_{m}) (R^{-1}_{nr}{P_{r} \over M})
  R_{ij} \int d \vec \xi \;\xi_{j}\;f_{k}(\vec \xi)\;\psi_{n}(\vec \xi)
   \label{boostend} \\ & & -{1 \over 2}(\Omega^{-1}_{kl}R^{-1}_{lm}S_{m})
		   (\Omega^{-1}_{ns}R^{-1}_{sr}S_{r})
     R_{ij} \int d \vec \xi \;\xi_{j}\;f_{k}(\vec \xi)\;f_{n}(\vec \xi)
\nonumber
\\
   & & -{1 \over 2}(R^{-1}_{kl}{P_{l} \over M})
		   (R^{-1}_{nm}{P_{m} \over M})
  R_{ij} \int d \vec \xi \;\xi_{j}\;\psi_{k}(\vec \xi)\;\psi_{n}(\vec
\xi).  \nonumber \endeqnarray

Now let us take into account, that when in the initial decomposition
(5) the classical solution is taken as a static soliton $u(\vec \xi)$, the
resulting theory turns out to be essentally non-relativistic, what results,
 in particular, in the lowest-order soliton Hamiltonian of the form $$
H=M+P^2/2M + \hbox { \rm quantum corrections}.$$ So by equating the
soliton boost operator (\ref{boostend}) to corresponding operator of the
spinning particle given by eq.(\ref{boostrep}) we have to use in the
latter the non-relativistic approximation too and replace $P^0$ by $M$ to
the leading order. Then the final result is the following set of
subsidiary conditions imposed on $u(\vec x)$ \eqnarray \int d \vec \xi \;
\xi_{i} \; \psi_{j}(\vec \xi) \; \psi_{k}(\vec \xi) & = & 0 ,
\label{newpp} \\ \int d \vec \xi \; \xi_{i} \;f_{j}(\vec \xi) \;
     f_{k}(\vec \xi) & = & 0 , \label{newff} \\ \int d \vec \xi \;
      \xi_{i}\; \psi_{j}(\vec \xi)\;f_{k}(\vec \xi) & = & {1 \over
       2}\varepsilon_{ijl}\;\Omega_{lk} .  \label{newpf} \endeqnarray

These relations can be understood as a criterion of
"particle-likeness" for the classical soliton field,
describing a spinning particle. It should be noted, that whereas the
orthogonality conditions (16--18) are valid for any static classical
solution due to the general properties of lorentz-covariance and so are
automatically consistent with equations of
motion [9], the relations (26--28) are more strong and restrictive. Namely,
the eqs. (17) and (18) are the direct consequences from eqs.(26) and (28)
correspondingly. Moreover, there might exist a static solution $u(\vec
x)$, that describes a two-soliton configuration and so cannot be
consistent with the one-particle representation of the Poincare group. In
this case the relations (26--28) obviously do not hold.

On account of these general considerations we show now that the typical
hedgehog configurations of nonlinear $\sigma$-models  describe
spinning particles independently of the profile of their chiral angles.
In two spatial dimensions,  we consider the
$O(3) \ \sigma$-model, described  by the Lagrangean density \equation {\cal
L} = {1 \over 2} \partial_{\mu} \varphi^{a} \partial^{\mu} \varphi^{a}
\label{lagro3} \endequation with subsidiary condition \equation
\varphi^{a} \varphi^{a} = 1.  \endequation This theory is a planar analog
of the Skyrme model and is hoped to reveal the fractional
spin and statistics after adding the Hopf term \cite{anyone,IJMP}.  The
standard
one-particle solution of the model is given by the
"baby-hedgehog" $Ansatz$ \cite{Belavin} \equation \varphi^{1} = \phi (r)
\cos n \vartheta , \; \;\; \varphi^{2} = \phi (r) \sin n \vartheta , \; \;
\; \varphi^{3} = { (1- \phi^{2} ) }^{1/2}, \label{solo3} \endequation
where $r,\vartheta$ are polar coordinates and \equation \phi (r) = {
4r^{n} \over r^{2n}+4 }\;, \label{phio3} \endequation and describes the
''baby-skyrmeon'' configuration with the topological charge $Q=n$ and the
mass $M=4 \pi Q$.  In the case of 2 spatial dimensions we have two
translational $ \psi^{a}_{i}=\partial_{i}\varphi^{a}, \;\; (i=1,\; 2) $
and only one rotational $ \;
f^{a}=\varepsilon_{ij}\;\xi_{i}\partial_{j}\varphi^{a} \; $ zero modes for
each isospin component  $\; \varphi^{a} \;$. The relations
(\ref{newpp}--\ref{newpf}) take now the form \eqnarray \int d \vec \xi
\; \xi_{i} \; \psi^{a}_{j}(\vec \xi) \; \psi^{a}_{k}(\vec \xi) & = & 0 ,
\label{newpp2} \\ \int d \vec \xi \; \xi_{i} \;   f^{a}(\vec \xi)
f^{a}(\vec \xi)  & = & 0 , \label{newff2} \\ \int d \vec \xi
\; \xi_{i}\; \psi^{a}_{j}(\vec \xi)\;f^{a}(\vec \xi) & = & {1 \over
     2}\varepsilon_{ij}\;\Omega \ , \label{newpf2} \endeqnarray where
      \equation \Omega=\int d \vec \xi \;  f^{a}(\vec \xi) f^{a}(\vec \xi)
=2 \pi n^2 \int \limits_0^{\infty} r dr \ \phi^2(r). \label{omega}
   \endequation It is easy to verify, that the
   straightforward substitution of eqs.(\ref{solo3}) into the relations
(\ref{newpp2}--\ref{newpf2}) leads to eq.(\ref{omega}) as the consistency
condition. Thus, the baby-skyrmeon solution (31) corresponds to the
spinning particle for any choice of the chiral angle (32).

In 3+1-dimensional spacetime, we consider the $SU(2)$-Skyrme model
  described by the Lagrangean \cite{skyrmeon}
\equation
  {\cal L} = -{1 \over 4 } \; tr \; L^{2}_{\mu} +
             { 1 \over 32} \; tr \; { [ L_{\mu} L_{\nu} ] }^{2},
\label{lagrsk}
\endequation
where, as usually, $\;L_{\mu}=U^{-1}\partial_{\mu}U\;$ is
the left chiral current and $\; U=\exp \{ i \tau_{a} \phi_{a} \}\;$
is the chiral field ( $ \tau_{a}\; $ are Pauli matrices).
In terms of three independent fields $ \phi_{a} $
the expression (\ref{lagrsk}) can be rewritten as \cite{Cebula}
\equation
   {\cal L} = {1 \over 2} \; \dot \phi_{a} \; M_{ab}(\vec \phi) \;
                         \dot \phi_{b} - V(\vec \phi),
\label{lagrskM}
\endequation
where
\eqnarray
   M_{ab}(\vec \phi) & = & {\phi_{a}\phi_{b} \over {|\phi|}^{2}}
            + {{\sin}^{2}|\phi| \over {|\phi|}^{2}}
	  \left( \delta_{ab}-{\phi_{a}\phi_{b} \over {|\phi|}^{2}} \right)
       +  \left( {{\sin}^{2}|\phi| \over {|\phi|}^{4}} -
		   {{\sin}^{4}|\phi| \over {|\phi|}^{6}}  \right) \; \times
\nonumber
\\
 & \times & \left[ (\partial_{i}\phi_{c}\partial_{i}\phi_{c})\phi_{a}\phi_{b}
	 +{1 \over 2} \partial_{i}{|\phi|}^{2}
	   \left( {1 \over 2} \partial_{i}{|\phi|}^{2}\delta_{ab}
		- \partial_{i}(\phi_{a}\phi_{b} \right) \right]
\label{M}
\\
     & + & {{\sin}^{4}|\phi| \over {|\phi|}^{4}}
	  \left[ (\partial_{i}\phi_{c}\partial_{i}\phi_{c})\delta_{ab}
	       - \partial_{i}\phi_{a}\partial_{i}\phi_{b} \right] ,
\nonumber
\endeqnarray
\equation
   V(\vec \phi)={1 \over 2} \left[ \gamma_{ii} + {1 \over 2}
       ( \gamma_{ii}\gamma_{jj} - \gamma_{ij}\gamma_{ij}) \right] ,
\label{V}
\endequation
and $\; \gamma_{ij}\; $ is given by
$$
    \gamma_{ij}={1 \over 4 {|\phi|}^{2}}
	  \left( 1- { {\sin}^{2} |\phi| \over {|\phi|}^{2}} \right)
	  \partial_{i}{|\phi|}^{2}\partial_{j}{|\phi|}^{2}
	+  { {\sin}^{2}|\phi| \over {|\phi|}^{2}}
	     (\partial_{i}\phi_{c}\partial_{j}\phi_{c} ) .
$$
{}From the Lagrangean (\ref{lagrskM}) we find the Hamiltonian  density
\equation
   {\cal H} = {1 \over 2} \; \pi_{a} \; M^{-1}_{ab}(\vec \phi) \; \pi_{b}
    + V(\vec \phi),
\label{hamskM}
\endequation
where the canonical
field momentum is $\; \pi_{a}=M_{ab}(\vec \phi)\;\dot \phi_{b} $.
(Since the Skyrme model is non-renormalizable, our
treatment of the model should be essentially quasiclassical and so
doesn't take care of the operator ordering.)

The treatment of the Skyrme model differs from the theories
considered below in that point, that it contains terms of the  4th order in
derivatives. As a result, all the scalar products of the model, in
particular, the orthogonality conditions (\ref{ortpp}--\ref{ortff}) in
the vicinity of the static classical solution $ \; \phi^{a}_{c}(\vec \xi)
\; $ acquire a nontrivial integration measure. It is easy to verify, that for
 the Lagrangean (\ref{lagrskM})
the weight function in the integration measure is
$ \; M_{ab}(\phi_{c}(\vec \xi))\; $.
Namely, the general form of the
orthogonality conditions for translational modes  can be written as [9] $$ \int
{ \partial
    {\cal L} (\vec \phi) \over \partial \partial^{j} \phi_{a}(\vec \xi)}
       \; \partial_{i} \phi_{a}(\vec \xi) \; d \vec \xi= M \delta_{ij}.
$$ The straightforward substitution of the Lagrangean (\ref{lagrskM})
in this expression gives
  $$ { \partial {\cal L} (\vec \phi) \over \partial
   \partial^{j} \phi_{a}(\vec
       \xi)} \; \partial_{i} \phi_{a}(\vec \xi)
   = \partial_{i} \phi_{a} \; M_{ab}(\vec \phi) \; \partial_{j} \phi_{b}
   \equiv \psi^{a}_{i} \; M_{ab}(\vec \phi) \; \psi^{b}_{j}.$$
With minor modifications this result can be extended to other relations
between translational and rotational zero modes, i.e. the
orthogonality  conditions take now the form
\eqnarray
   \int d \vec \xi \;
   \psi^{a}_{i}(\vec \xi) \; M_{ab}(\phi_{c}(\vec \xi))\; \psi^{b}_{j}(\vec
   \xi) & = & M\delta_{ij},
\label{ortpMp}
\\ \int d \vec
     \xi \; \psi^{a}_{i}(\vec \xi) \; M_{ab}(\phi_{c}(\vec \xi)) \;
      f^{b}_{j}(\vec \xi)  & = & 0,
\label{ortpMf}
\\
\int d \vec \xi \;
f^{a}_{i}(\vec \xi) \; M_{ab} (\phi_{c}(\vec \xi))\; f^{b}_{j}(\vec \xi)
& = & \Omega_{ij}  = \Omega_{i}\delta_{ij}.
\label{ortfMf}
\endeqnarray

In a more general approach, the weight function for the scalar
product of zero modes in the Skyrme model can be determined through the
small fluctuations equation, although the complete  treatment of the
fluctuation spectrum should be ambiguous due to non-renormalizabilty of
the model.  The corresponding linearized equation, in the neighborhood of
the static classical solution $ \vec \phi_{c}(\vec x) $, takes the form
\equation M_{ab}(\phi_c (\vec x)) \ \partial_{t}^2 \Phi_b(\vec x, t)
=\Lambda_{ab}(\phi_c(\vec x), \vec \partial) \ \Phi_b(\vec x, t),
\label{fluctuations} \endequation where $\Lambda_{ab}=\Lambda_{ba}^+$.
{}From the latter property of $\Lambda_{ab}$ it is easy to
verify by conventional methods, that the scalar product for
eigenfunctions of the corresponding spectral problem should be taken indeed in
the form
$$ \int \ (\psi_1^a)^* M_{ab}(\phi_c) \psi_2^b \ .$$

Now, using the same reasoning as by derivation of  relations
(\ref{newpp}--\ref{newpf}),
we get the following set of "particle-likeness" conditions for the
classical solutions in the Skyrme model
 \eqnarray \int d
    \vec \xi \; \xi_{i} \; \psi^{a}_{j}(\vec \xi) \; M_{ab}(\phi_{c}(\vec
    \xi)) \; \psi^{b}_{k}(\vec \xi) & = & 0 , \label{newpMp} \\ \int d
    \vec \xi \; \xi_{i} \;f^{a}_{j}(\vec \xi) \; M_{ab}(\phi_{c}(\vec
     \xi)) \; f^{b}_{k}(\vec \xi)  & = & 0 , \label{newfMf} \\ \int d \vec
      \xi \; \xi_{i}\; \psi^{a}_{j}(\vec \xi)\; M_{ab}(\phi_{c}(\vec
	  \xi)) \;  f^{b}_{k}(\vec \xi) & = & {1 \over
       2}\varepsilon_{ijl}\;\Omega_{lk} \ , \label{newpMf} \endeqnarray
where the moments of inertia of the classical configuration are
defined in (\ref{ortfMf}).
Now we verify the eqs.(46-48) for the standard hedgehog configuration
\equation
    \phi_{a}={r_{a} \over r}\phi(r)=n_{a}\phi(r).
\label{phisk}
\endequation
In this case
\equation
   M_{ab}(r)  =  \delta_{ab} \; f_{1}(r) + n_{a}n_{b} \; f_{2}(r),
\label{Mhedgehog}
\endequation
where $ \; f_{1}(r) \; $ and $ \; f_{2}(r) \; $ are given by
\eqnarray
  f_{1}(r) & = & {{\sin}^{2}\phi \over \phi^{2}}
       \left( 1+ {\phi'}^{2} + {{\sin}^{2}\phi \over r^{2}}  \right),
\nonumber
\\
  f_{2}(r) & = & 1 - {{\sin}^{2}\phi \over \phi^{2}}
      \left( 1 - {1 \over r^{2}} (2 \phi^{2} -{\sin}^{2}\phi)
	   + {\phi'}^{2} \right).
\nonumber
\endeqnarray
For this configuration the integrals in the l.h.s. of eqs.(\ref{newpMp})
and (\ref{newfMf}) vanish exactly provided  by the symmetry
properties of the $Ansatz$ (49). Further, after some algebra the l.h.s of
eq.(\ref{newpMf}) yields
\equation
\int d \vec \xi \; \xi_{i}\;
   \psi^{a}_{j}(\vec \xi)\; M_{ab} \; f^{b}_{k}(\vec \xi) = {4\pi
   \over 3} \varepsilon_{ijk} \; \int r^{2}  dr \; \phi^{2}(r)
         f_{1}(r),
\endequation
while
\equation
\Omega_{ij} = {8\pi \over
    3} \delta_{ij} \int r^{2} dr \; \phi^{2}(r)  f_{1}(r) .
\endequation
So we obtain, that the conditions of particle-likeness
(\ref{newpMp}--\ref{newpMf}) for the hedgehog configuration are
satisfied in 3 spatial dimensions as well, and once more it holds
independently of the profile of the function $\; \phi(r) \;$. Thus,
the soliton (\ref{phisk}) of the $SU(2)$-Skyrme model might be embedded into
the
irreductible representation of the Poincare group for the particle with
spin without any restrictions on the shape of the chiral angle.

To conclude let us mention, that the present analysis can be easily
extended to other soliton models including vector fields, etc.
On the other hand, the relations (26-28), being independent of
 equations of motion, can play an
essential role of additional constraints in approximate
calculations as well. For example, they can be explored as a test for
various sample functions, used in describing the shape of the skyrmeon
[1,5].  Concerning the Skyrme model, our analysis is consistent with the
well-known result [5,14], that the spin of $SU(2)$-skyrmeon can be
arbitrary. The results of application of the present analysis to the
$SU(3)$-skyrmeons with the half-odd spin  will be reported
separately.

This work is supported in part by Russian Universities Scientific
Program.

\vfill \eject

\normalsize
\rm

\end{document}